\documentclass[12pt]{article}
\usepackage{amsmath}
\usepackage{amssymb}
\usepackage{graphicx}

\newcommand{\vev}[1]{\left\langle#1\right\rangle}

\newcommand{\Lag}{ {\mathcal L} }

\newcommand{\Journal}[5]{#1~#2~\textbf{#3}, #5 (#4)}
\newcommand{\eff}{{\rm eff}}

\newcommand{\dpn}[1]{\frac{d^#1\!p}{(2\pi)^#1}}
\newcommand{\Tr}{\mathrm{Tr}}

\begin{document}

\title{Coexistence of color superconductivity and chiral symmetry breaking within the NJL model}
\author{D.Blaschke$^{1,2}$, M.K.Volkov$^2$, and V.L.Yudichev$^2$\\[5mm]
\itshape\small $^{1)}$Fachbereich Physik, Universit\"at Rostock,
D-18051 Rostock, Germany\\
\itshape\small$^{2)}$Joint Institute for Nuclear Research,
141980 Dubna, Russian Federation}
\date{}
\maketitle

\begin{abstract} The phase diagram for quark matter is investigated
within a simple Nambu--Jona-Lasinio model without vector
correlations. It is found that the phase structure in the
temperature--density plane depends sensitively on the
parametrization of the model. We present two schemes of
parametrization of the model where within the first one a first
order phase transition from a phase with broken chiral symmetry to
a color superconducting phase for temperatures below the triple
point at $T_t= 55$ MeV occurs whereas for the second one a second
order phase transition for temperatures below $T_t = 7$ MeV is
found. In the latter case, there is also a coexistence phase of
broken chiral symmetry with color superconductivity, which is a new
finding within this class of models. Possible consequences for the
phenomenology of the QCD phase transition at high baryon densities
are discussed.
\end{abstract}
\noindent{Rostock University\\ Preprint No. MPG-VT-UR 235/02}

\clearpage
\section{Introduction}
The phenomenon of color superconductivity
\cite{%
barrios77,bailin84,iwasaki:plb95,iwasaki:ptp95,alford98,%
rapp98,halasz98,Stephanov:1998dy,bloch,alford:npb99,pisarski99,son99,schaefer99,berges99,%
rapp00,pisarski00,rajagopal00,alford:arnps01,alford:prd01,buballa02,%
oertel02,gastineau02,harada02,steiner02,neumann02,bowers02%
}
is of general interest,
in particular, in  studies of the QCD phase structure
\cite{bloch,alford:npb99,berges99,rapp00,rajagopal00,alford:arnps01,alford:prd01,buballa02,oertel02,%
gastineau02,harada02}
and applications in the astrophysics of compact stars \cite{alford:arnps01,lnp}.
Observable consequences are expected for, e.~g., the cooling behavior \cite{bkv,bgv}.
Different aspects have been investigated so far,
whereby  models of the NJL type have been
widely employed \cite{ebert86,volk86,ebert93,blaschke01} in studies of the
phase structure in the vicinity of
the hadronization transition.

Recently, it has been shown in these investigations that for
low temperatures ($T$) and not too large chemical potentials ($\mu$)
the two-Flavor Color Superconductivity (2SC) phase is favored
over alternative color superconducting phases
\cite{buballa02,oertel02,steiner02,neumann02}.
According to \cite{neumann02}, the Color-Flavor-Locked (CFL) phase
occurs only at $\mu \gtrsim 430$ MeV.

It is generally agreed that at low temperatures the
transition of the matter from the phase with broken chiral symmetry
to the color superconducting phase is of the first order
(see e. g. \cite{berges99}). From the point of view of
phenomenological applications, as e.g. in
compact star physics, the order of the phase transition to quark
superconducting matter plays an important role.
The conclusion about the first order phase transition was drawn
within models without vector interaction channels taken into account;
the vector interaction has been considered in few papers
\cite{langfeld99,buballa:prd01}.
It was found that the presence of quark interaction in the vector channel
moves the critical line in the in $\mu-T$ plain to larger $\mu$
\cite{berges99,klimt90,asakawa89}.
Recently it has been demonstrated \cite{kitazawa} that the
critical line of  first order phase transition
in the $\mu-T$ plane can have a second end-point at low temperatures,
besides the well known one at high temperatures.
The latter one could even be subject to experimental verification in
heavy-ion collisions \cite{Stephanov:1998dy} whereas the former could be of relevance
for neutron stars.
While in Ref.~\cite{kitazawa} this feature of the phase diagram was a consequence of
the presence of interaction in the vector channel, we would like to investigate
in the present work
the sensitivity of the phase diagram to the choice of model parameters without
interaction in the vector channel.
We will demonstrate that in the absence of the vector channel interaction the
phase transition is not necessarily of the first order, thus revising
statements in Refs.~\cite{asakawa89,kunihiro91}.

It is worth noting that some progress has recently been done in lattice calculations.
There are methods being developed that allow to extend lattice results to the case of
finite chemical potentials \cite{karsch,fodor,fodor1}. However, these methods are
valid only for small chemical potentials (see e.~g.~\cite{karsch}), below
the conditions at which the color superconductivity phase is expected to form.

The structure of our paper is as follows.
In Sect.~2, a chiral quark model is introduced, its Lagrangian is given
and the model parameters are fixed from the vacuum state in two different
schemes.
Temperature and chemical potential are introduced into the
quark model in Sect.~3, using the Matsubara formalism.
The conclusions and a discussion of the obtained results are given in Sect.~4.

\section{NJL model with the scalar diquark channel}
In order to study the quark matter phase diagram including color superconductivity,
one should generalize the concept of the single order parameter related to the
quark-antiquark condensate in the case of chiral symmetry breaking to a set of
order parameters when condensation can occur in other interaction channels too.
The simplest extension is the scalar diquark condensate $\vev{\psi\psi}$ for $u$
and $d$ quarks
\begin{equation}
\label{delta}
\delta=\vev{\bar\psi i\gamma_5\tau_2\lambda_2C\bar\psi^T},
\end{equation}
which is an order parameter characterizing the domain where the color symmetry is
spontaneously broken and the quark matter finds itself in
the (two-flavor) color superconducting (2SC) state.
This quantity is the most important one among other possible
condensates that can be constructed in accordance with
the Pauli principle \cite{neumann02}.
In (\ref{delta}) the matrix $C$ is the charge conjugation
matrix operator for fermions
\begin{equation}
C=i\gamma_0\gamma_2.
\end{equation}
The matrices $\tau_2$ and $\lambda_2$ are
Pauli and Gell-Mann matrices, respectively.
The first one acts on the flavor indices of spinors while the
second one acts in the color space.

If the electroweak interaction is discarded and only the strong coupling
is in focus, the resulting quark matter phase diagram is
essentially determined by nonperturbative features of the QCD vacuum state.
One therefore has  to resort to nonperturbative approaches
to describe the behavior of particles at various conditions, ranging
from cold and dilute  matter up to the hot and dense one.
A reliable and widely tested model to nonperturbative strong coupling QCD
is provided by the Dyson-Schwinger equations \cite{Roberts:2000aa},
however, for qualitative studies like the one we attempt here it proves to be
too complex.
Therefore, we will use here a simple and tractable nonperturbative model of
quark interaction, the Nambu--Jona-Lasinio (NJL) model
\cite{ebert86,volk86,ebert93,blaschke01,klevansky92,hatsuda94}, which
 has been extensively exploited for the description of the properties
of the light meson sector of QCD (also to describe the
color superconductivity phase \cite{buballa02,ebert,schwartz99}) and proved to be
a model respecting the low-energy theorems.
Before we proceed to the case of finite temperature and density,
the model parameters that determine the
quark interaction should be fixed. This shall
be done for the vacuum state where hadronic properties are known.
We will assume, according to common wisdom that, once fixed,
these parameters (originating from the nonperturbative gluon
sector of QCD) will not change, even in the vicinity of the transition
to the quark matter.
This transition is thus caused by medium effects in
the quark sector only.

\subsection{Lagrangian}
In the present paper we restrict ourselves to the two-flavor case,
leaving the strange quark and effects related to it
beyond our consideration.
As we constrain ourselves to only two order parameters, the quark and scalar diquark condensates,
during our investigation, the interaction of
quarks  will be represented in the Lagrangian by $SU(2)_L\times SU(2)_R$
symmetric scalar, pseudoscalar quark-antiquark, and scalar diquark vertices:
\begin{eqnarray}
\Lag&=&\bar\psi(i\!\!\not\!\partial-\hat m^0)\psi+\Lag_{q\bar
q}+\Lag_{qq},\\
\Lag_{q\bar q}&=&\frac{G}{2}\left[(\bar\psi\psi)^2+(\bar\psi
i\gamma_5\vec{\tau}\psi)^2\right],\\
\Lag_{qq}&=&\frac{H}{2}(\bar\psi
i\gamma_5\tau_2\lambda_2C\bar\psi^T)(\psi^{T} C
i\gamma_5\tau_2\lambda_2\psi),
\end{eqnarray}
where $\hat m^0$ is the diagonal current quark mass matrix
$\hat m^0=\mathrm{diag}(m_u,m_d)$,
$G$ and $H$ are  constants describing the interaction of quarks in
the scalar, pseudoscalar, and scalar diquark channels, respectively.
We work in the isospin symmetric case $(m_u^0=m_d^0\equiv m^0)$, thus
$\hat m^0= m^0 \mathbf 1_f$.

By the standard Hubbard-Stratonovich procedure , we
introduce auxiliary scalar ($\sigma$), pseudoscalar triplet ($\vec{\pi}$),
and diquark ($\Delta, \Delta^\ast$) fields together with
Yukawa-like terms in the Lagrangian density instead of the four-quark
vertices:
\begin{eqnarray}\label{Lag1}
&& \tilde \Lag=\bar\psi(i\!\not\!\partial-m^0+\sigma+i\gamma_5\vec{\tau}\vec{\pi})\psi-\nonumber\\
&&\qquad-\frac12 \Delta^{\ast} \psi^T C\gamma_5\tau_2\lambda_2\psi+
\frac12\Delta\bar\psi\gamma_5\tau_2\lambda_2C\bar\psi^T\nonumber\\
&&\qquad-\frac{\sigma^2+\vec{\pi}^2}{2G}-\frac{|\Delta|^2}{2H}~.
\end{eqnarray}
In order to integrate out the quark degrees of freedom by Gaussian path integration,
it is appropriate to represent the quark fields by the bispinor
\begin{equation}
q(x)=\left({\psi(x) \atop C\bar\psi^T(x)}\right),
\end{equation}
and to introduce the matrix propagator $S(p)$:
\begin{equation}\label{invpropagator}
S^{-1}(p)=
\left(
\begin{array}{cc}
\not\!p - \hat M &
\Delta\gamma_5\tau_2\lambda_2\\
-\Delta^\ast\gamma_5\tau_2\lambda_2&
\not\!p - \hat M
\end{array}
\right).
\end{equation}
Integrating over $q(x)$ and $\bar q(x)$, we then obtain an effective
Lagrangian in terms of collective scalar and pseudoscalar quark-antiquark
and scalar diquark excitations. Here, we restrict ourselves to
the mean-field approximation, leaving the next-to-leading order
corrections in the $1/N_c$ expansion beyond our model.
Finally, the effective Lagrangian density reads
\begin{equation}
{\cal L}_{\rm eff}=-\frac{\sigma^2+\vec{\pi}^2}{2G}-\frac{|\Delta|^2}{2H}
 -i\!\int\!\dpn{4}\frac12 \Tr\ln\left(S^{-1}(p)\right).
\end{equation}
The trace in (\ref{invpropagator}) is taken in the Dirac, color, and
flavor space.
The matrix $\hat M$ contains $\sigma$ and $\vec{\pi}$ fields:
\begin{equation}\label{M}
\hat M=(m^0-\sigma) \mathbf{1}-i\gamma_5\tau_a\pi_a,
\end{equation}
the sum over $a=1,2,3$ is assumed,
and $\mathbf{1}=\mathbf{1}_c\cdot \mathbf{1}_f\cdot \mathbf{1}_D$.

As it was mentioned above, we are working in the mean-field approximation
and the quark condensates are of interest.
Therefore, further study can be performed
in terms of the effective potential
\begin{equation}\label{Veff}
  V_\eff=-\lim_{v_4\to\infty}\frac{1}{v_4}\int_{v_4}\! d^4x \Lag_\eff
\end{equation}
where $v_4$ is  4-dimensional volume.
The vacuum expectation values of the collective variables
$\sigma$, $\vec{\pi}$, $\Delta$, and $\Delta^\ast$ determine the
absolute minimum of $V_\eff$. They are given by the equation
\begin{equation}\label{Vmin}
\frac{\partial V_{\rm eff}}{\partial \sigma}=\frac{\partial
  V_{\rm eff}}{\partial \Delta}=\frac{\partial V_{\rm eff}}{\partial \Delta^\ast}=0~.
\end{equation}
\textit{A priori} it is known that
in the vacuum only the $\sigma$ field acquires a nonvanishing expectation
value. The diquark fields  $\Delta$, $\Delta^\ast$ are expected
to have nonzero mean values only in dense matter.
The mean value of the pseudoscalar isotriplet field  $\vec{\pi}$
is always equal to zero, therefore we omit it hereafter.

Having solved Eq.~(\ref{Vmin}) for the field $\sigma$, one can
work in terms of the constituent quark mass $m$, connected with the
current quark mass by the gap equation
\begin{equation}\label{gap}
m_0-m=\vev{\sigma}=2 G\langle\bar\psi\psi\rangle~.
\end{equation}
In the chiral limit ($m^0=0$), the constituent quark mass is
proportional to the quark condensate and thus can be treated as the order
parameter.

In the NJL model the quark condensate is
\begin{equation}
\langle\bar\psi\psi\rangle=-4 m I_1^{\Lambda}(m),
\end{equation}
where
\begin{equation}
I_1^{\Lambda}(m)=\frac{-i N_c}{(2\pi)^4}\int \!\!\theta(\Lambda^2-\vec{p}^2)
\frac{d^4p}{m^2-p^2}~.
\end{equation}
The divergence in $I_1^{\Lambda}(m)$ is eliminated by means of
  a sharp 3D cut-off at the scale $\Lambda$.

\subsection{Parameter fixing}
In our model we have four parameters: the
four-quark interaction constants $G$ and  $H$, cut-off
$\Lambda$,   and the current quark mass $m^0$.
Without diquarks, there are only three: $G$, $\Lambda$, and
$m^0$.  They are fixed by the following relations:
\begin{enumerate}
\item The Goldberger-Treiman relation (GTR):
\begin{equation}
m=g_\pi F_\pi~,
\end{equation}
where $F^{\rm exp}_\pi\approx 93$ MeV is the pion weak coupling
constant and $g_\pi$ describes the coupling of a pion with quarks
$g_\pi\vec{\pi}\bar\psi\vec{\tau}\psi$
\begin{equation}
g_\pi^{-2}=4 I_2^{\Lambda}(m),\quad
I_2^{\Lambda}(m)=\frac{-iN_c}{(2\pi)^2}
\int\!\frac{\theta(\Lambda^2-\vec{p}^2)d^4p}{(m^2-p^2)^2}~.
\end{equation}
\item[2-a.] The quark condensate (QC) from QCD sum rules
\begin{equation}
\langle\bar\psi\psi\rangle_{\rm QCDSR}=
-4 m I_1^{\Lambda}(m)\approx (-240\ \mbox{MeV})^3.
\end{equation}
\item[2-b.] The decay constant $g_\rho$ for the $\rho\to 2\pi$ (R2PD) process
\begin{equation}
g_\rho=\sqrt{6}g_\pi,  \quad
g_\rho^{\rm exp}\approx 6.1 ~.
\end{equation}
The $\pi-a_1$ transitions are omitted here.
\item[3.] The current quark mass $m^0$ is fixed from the GMOR relation:
\begin{equation}
M_\pi^2=\frac{-2 m^0\vev{\bar\psi\psi} }{ F_\pi^2}, \quad M_\pi^{\rm exp}\approx 140\;
\mbox{MeV}~.
\end{equation}
In the chiral limit $M_\pi=0$, $m^0=0$.
\item[4.] With the diquark channel included, there is an additional parameter
$H$ which can  be fixed as
$H=3/4 G$ from the Fierz transformation (as e.~g.{} in\cite{buballa02})
\footnote{
Some authors use $H=1/2G$. It turned out that within our
model the resulting phase diagram is not much affected if
one makes the choice in favor of $H=1/2G$.
However, it would be preferable to fix the constant $H$
from some observable, e.~g.{} from the nucleon mass.
}.
\end{enumerate}

In the item 2, we have given two alternatives: one can either use the value of the
quark condensate taken from QCD sum rule estimates or demand from the model
that it should describe the $\rho\to 2\pi$ decay. The latter
is well observable in experiment contrary to the quark condensate.

For simplicity, we perform all calculations in the chiral limit $m^0=0$.
In this case, when investigating the hot and dense quark matter,
the borders between phases turn out to be sharp and the critical
temperature and chemical potential are well defined.
With the finite current quark mass, the transitions from
one phase to the other become smooth.

As a result, one obtains two different parameter sets shown in
Table~\ref{paramset}.

In the Type I parameter set the interaction of quarks is stronger,
the UV cut-off is smaller, and the constituent quark mass is greater.
One can calculate the dimensionless constant $G\Lambda^2$. It equals
4.6 for the Type~I and 3.72 for the Type II, respectively.
As we will see further, these two parametrizations result in qualitatively
different phase diagrams.

\section{NJL model at finite $T$ and $\mu$ }
\subsection{Thermodynamical potential}
We extend the NJL model to the case of finite  temperatures $T$ and
chemical potentials $\mu$,
applying the Matsubara formalism, and restrict ourselves to the isospin symmetric case
where up and down quark chemical potentials coincide.
The thermodynamical potential per volume is
\begin{eqnarray}\label{thpot}
\Omega(T,\mu)&=&-T\sum_n\!\int\! \dpn{3}\; \frac12 {\rm
  Tr}\ln\left(\frac{1}{T}\tilde S^{-1}(i\omega_n,\vec{p})\right)\nonumber\\
&+&\frac{\sigma^2}{2G}+\frac{|\Delta|^2}{2H}~,
\end{eqnarray}
where $\omega_n=(2n+1)\pi T$ are  Matsubara frequencies for fermions,
and the chemical potential is included into the definition of
inverse quark propagator
\begin{equation}
\tilde S^{-1}(p_0,\vec{p})=
\left(
\begin{array}{cc}
\not\!p - \hat M -\mu\gamma_0&
\Delta\gamma_5\tau_2\lambda_2\\
-\Delta^\ast\gamma_5\tau_2\lambda_2&
\not\!p - \hat M+\mu\gamma_0
\end{array}
\right).
\end{equation}
The expression in (\ref{thpot})
can be simplified using the equations
\begin{eqnarray}
\frac12 \Tr\ln\left(\tilde S^{-1}(i\omega_n,\vec{p})\right)
&=& 4\left[\ln\left(\frac{(\omega_n^2+{E^+}^2)(\omega_n^2+{E^-}^2)}{T^4}\right)\right]\nonumber\\
&+&2\left[\ln\left(\frac{(\omega_n^2+{\epsilon^+}^2)(\omega_n^2+
{\epsilon^-}^2)}{T^4}\right)\right]
\end{eqnarray}
and
\begin{equation}
T\sum_{n=-\infty}^{\infty}\ln\left(\frac{(\omega_n^2+{E^\pm}^2)}{T^2}\right)
=E^\pm+2 T\ln[1+\exp(-E^\pm/T)].
\end{equation}

\begin{eqnarray}\label{therm_potential}
\Omega(T,\mu)&=&-\int\dpn{3}\left\{2\left(2\epsilon\theta(\Lambda^2-\vec{p}^2) +
2 T \ln\left[1+\exp\left(-\frac{\epsilon^+}{T}\right)\right]\right.\right.\nonumber\\
&+&\left.2 T \ln\left[1+\exp\left(-\frac{\epsilon^-}{T}\right)\right]\right)\nonumber\\
&+&4\left(\left(E^++E^-\right)\theta(\Lambda^2-\vec{p}^2)+2 T \ln\left[1+\exp\left(-\frac{E^+}{T}\right)\right]\right.\nonumber\\
&+&\left.\left.2 T
	\ln\left[1+\exp\left(-\frac{E^-}{T}\right)\right]\right)\right\}
 +\frac{m^2}{2G}+\frac{|\Delta|^2}{2 H}~,
\end{eqnarray}
where
\begin{equation}
\epsilon=\sqrt{\vec{p}^2+m^2},\quad \epsilon^\pm=\epsilon\pm \mu,
\end{equation}
\begin{equation}
E^\pm=\sqrt{(\epsilon^\pm)^2+|\Delta|^2}.
\end{equation}
The cold matter limit $T=0$ looks as follows:
\begin{eqnarray}\label{coldmatterlimit}
\Omega(0, \mu)&=&-\!\int\!\dpn{3}\left[2(|\epsilon^+| +|\epsilon^-|)+
4(E^++E^-)\right]
\nonumber\\
&\times&\theta(\Lambda^2-\vec{p}^2)+\frac{m^2}{2G}+\frac{|\Delta|^2}{2 H}~.
\end{eqnarray}
The thermodynamical potential cannot be calculated in
closed form for arbitrary $T$ and $\mu$. However,
in the cold matter limit one can easily obtain analytic
expressions for the thermodynamical potential or
its derivatives if only one of the collective variables $\sigma$
or $|\Delta|$ has a nonvanishing  average value. This allows to
find what kind of phase transition is to be expected
for different parameter choices.

We evaluate the remaining 3D momentum integrals numerically and
calculate the value of thermodynamical potential at different
$T$ and $\mu$ for the two types of  model parameter sets.
The equilibrium state for each $T$ and $\mu$ is
determined by  $\vev{\sigma}=-m$ and $\vev{|\Delta|}$
corresponding to the minimum of $\Omega(T,\mu)$.

\subsection{Numerical results: Type I}
\begin{figure}[h]
\includegraphics[scale=1.0]{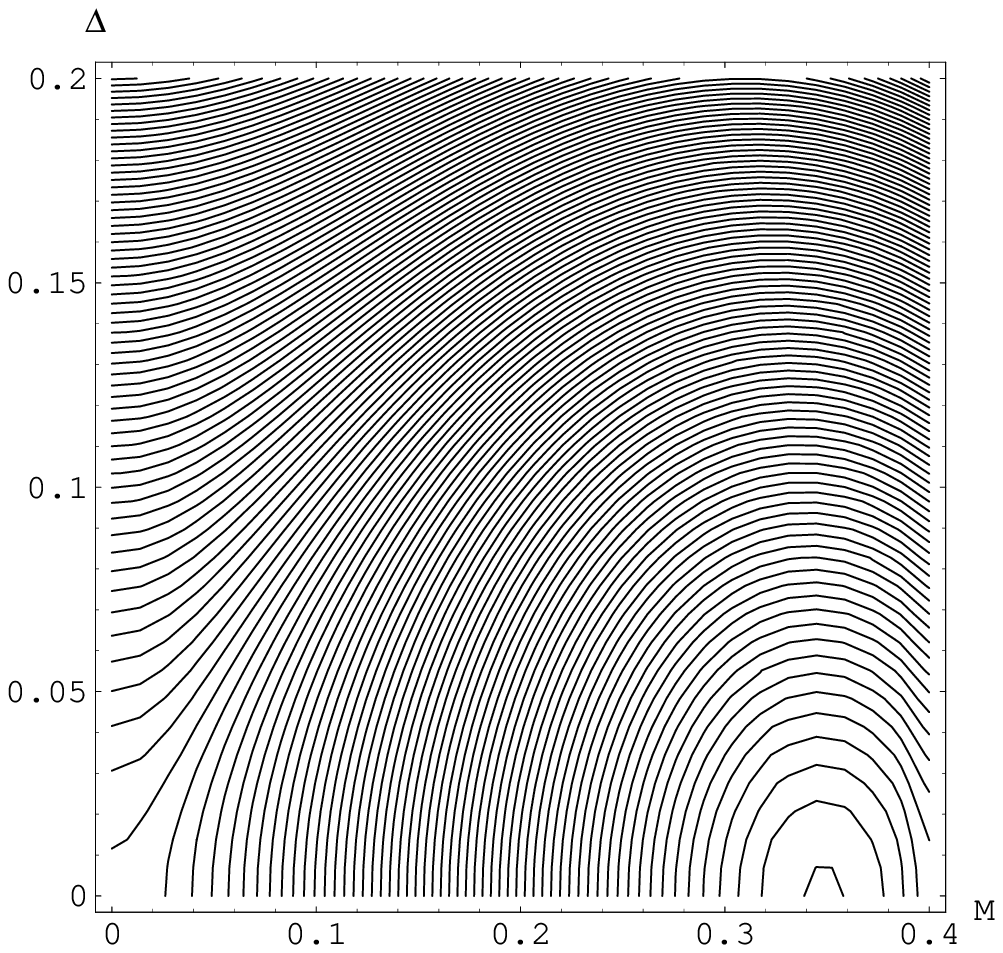}
\caption{The contour plot for the thermodynamic potential
as a function of of $m$ and $\Delta$ at zero temperature
and the chemical potential $\mu=0$ MeV.}
\label{thdp_I_1}
\end{figure}

\begin{figure}[h]
\includegraphics[scale=1.0]{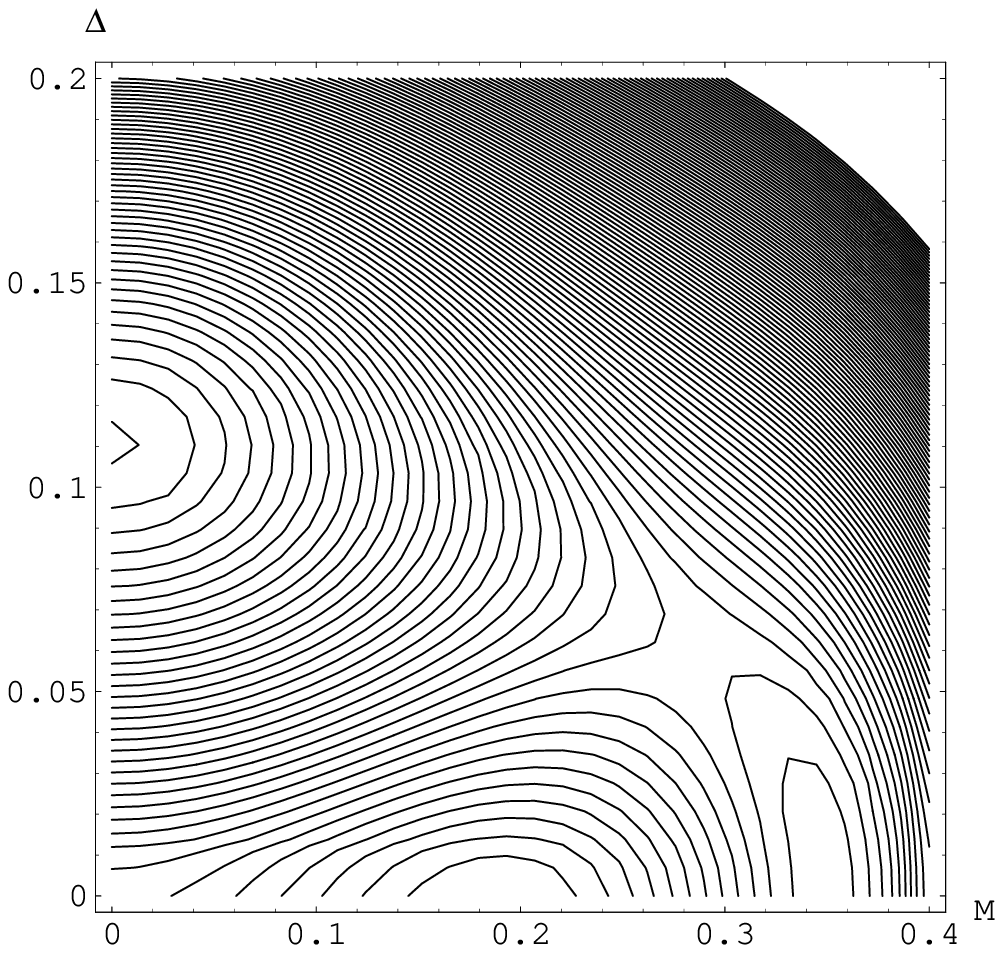}
\caption{The contour plot for the thermodynamic potential
as a function of of $m$ and $\Delta$ at zero temperature
and the chemical potential $\mu=$ 340 MeV.}
\label{thdp_I_2}
\end{figure}

\begin{figure}[h]
\includegraphics[scale=1.0]{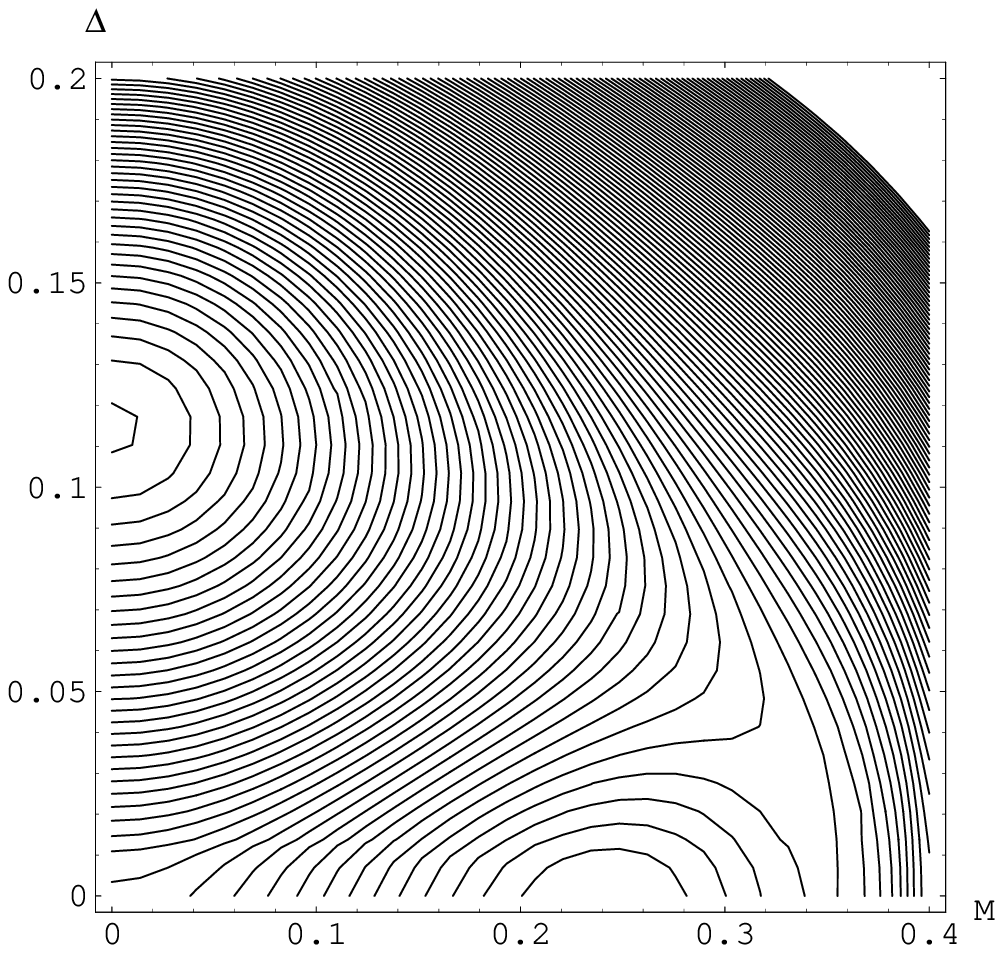}
\caption{The contour plot for the thermodynamic potential
as a function of of $m$ and $\Delta$ at zero temperature
and the chemical potential $\mu=$350 MeV.}
\label{thdp_I_3}
\end{figure}

\begin{figure}[h]
\includegraphics[scale=1.0]{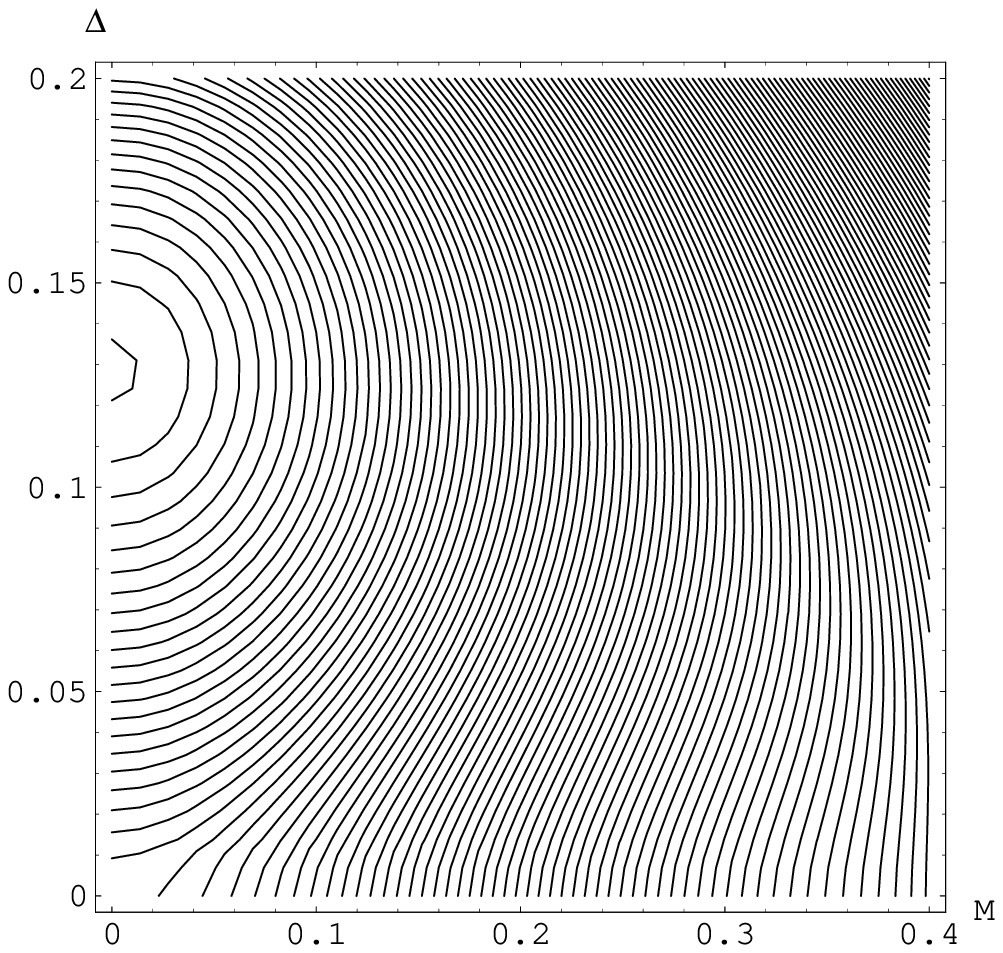}
\caption{The contour plot for the thermodynamic potential
as a function of of $m$ and $\Delta$ at zero temperature
and the chemical potential $\mu=$ 400 MeV.}
\label{thdp_I_4}
\end{figure}

It is quite illustrative to look at
the contour plots of the thermodynamical potential. For several
values of $\mu$ at $T=0$ they are shown in Figs.~\ref{thdp_I_1}--\ref{thdp_I_4} where
one can follow the appearance and disappearance of local minima,
maxima, and saddle points of the thermodynamical potential with
increasing chemical potential.
For zero temperature and chemical potential we have,  as expected,
a nonzero constituent quark mass (quark condensate) corresponding to
the absolute minimum of the thermodynamical potential at $m\sim 350$ MeV and
$\vev{|\Delta|}=0$ in Fig.~\ref{thdp_I_1}.
At a certain chemical potential, a new local minimum
related to the diquark condensate near $\vev{|\Delta|}\sim$110 MeV and $m=0$
(Fig.~\ref{thdp_I_2}), but it does not yet give
the absolute minimum.
There is also a local maximum around $m\sim 200$ MeV and $\vev{|\Delta|}=0$.
As the matter becomes more dense, the
second minimum lowers until it becomes degenerate with
the first minimum while the average value of $\sigma$ (or $-m$)
remains almost unchanged (see Fig.~\ref{thdp_I_3}).
Above the corresponding (critical) chemical potential
$\mu_c\approx 321$ MeV, the second minimum becomes the absolute one and
a first order transition occurs, during which
$\vev{\sigma}$ discontinuously changes to zero while
the diquark condensate acquires nonzero value
breaking the color symmetry of the strong interaction.
This characterizes the color superconducting phase transition in quark matter.
Furthermore, the local minimum on the $m$ axis merges the saddle point
(see Fig.~\ref{thdp_I_4}) and, at still higher $\mu$, only the local minimum on the
$\Delta$ axis  near $|\Delta|\sim$ 130 MeV and $m=0$ remains.

\begin{figure}
\includegraphics[scale=0.7]{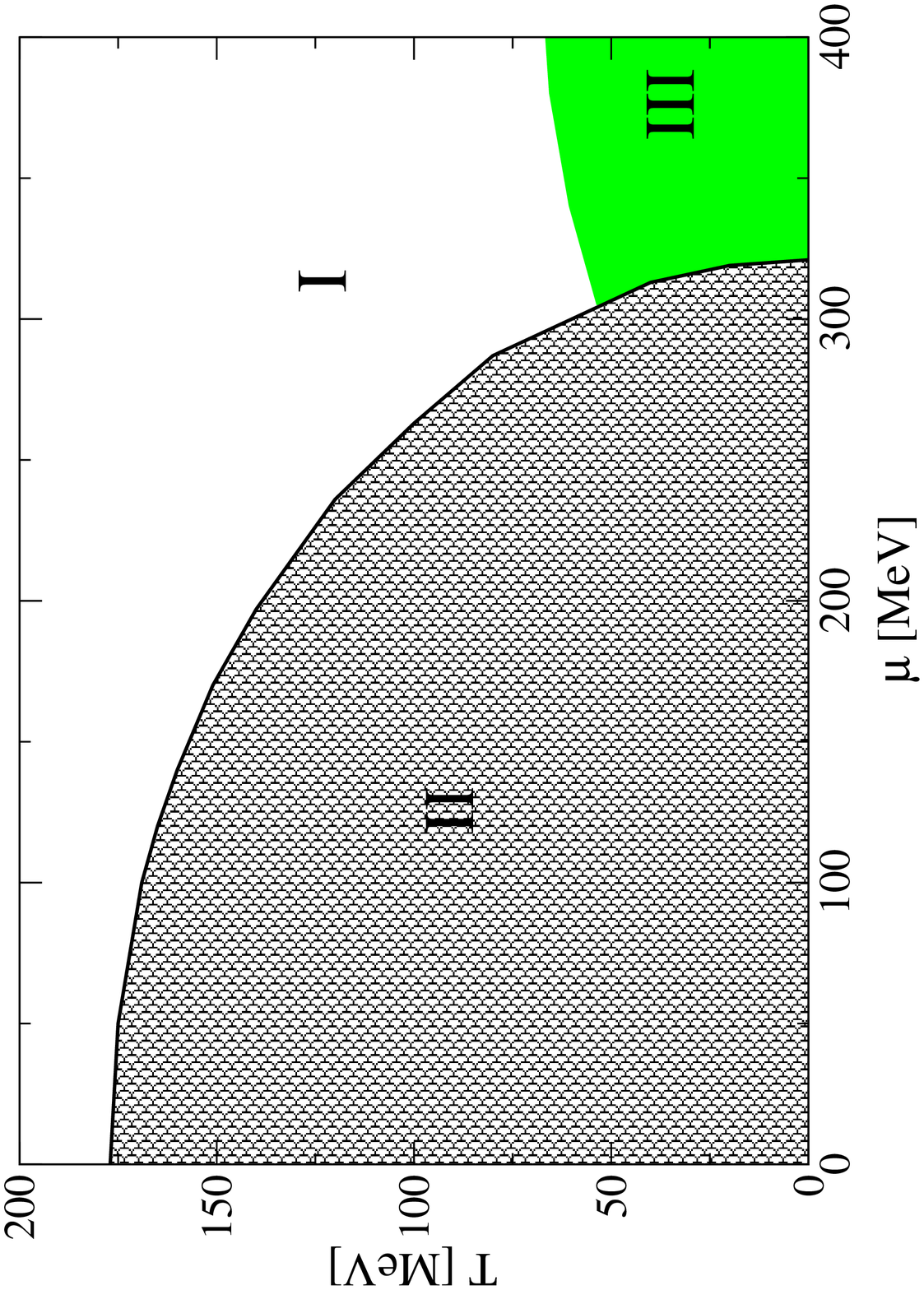}
\caption{The quark matter phase diagram from the NJL model
with the Type I parameter set. In phase I both chiral and diquark
condensates vanish; in phase II the chiral symmetry is broken; in
phase III the chiral symmetry is restored while the diquark condensate is nonzero.}
\label{phdI}
\end{figure}

At a fixed chemical potential above $\mu_c$, with the temperature
rising, the average value of $|\Delta|$ decreases until it reaches
zero at the critical temperature $T_c$ which can be
roughly estimated using the BCS theory formula
\begin{equation}
T_c\approx 0.57 \vev{|\Delta|}_{T=0}~.
\end{equation}
Above this temperature quark matter is in the symmetric phase
\footnote{According to recent investigations \cite{Kitazawa:2001ft}, a so-called
pseudo-gap phase as a precursor of color superconductivity can occur in this region.
}
where the chiral and color symmetries are restored.
Finally, we obtain the phase diagram shown on Fig.~\ref{phdI}
with three phases: the hadron phase, 2SC phase,
and symmetric phase. All three phases coexist at the triple point:
$T_{t}\approx 55$ MeV and $\mu_{t}\approx 305$ MeV.

\subsection{Numerical results: Type II}
As for the Type I  parameter set, at zero $T$ and $\mu$
only the constituent quark mass $m$, being the order parameter for the chiral
condensate, is nonzero, whereas the diquark gap $\Delta$ vanishes.
 However, the vacuum value of $m$ is
lower than that for the Type I and, with the chemical
potential increasing, $\mu$ becomes equal to the vacuum value of
$m$ before the second local minimum, corresponding to the diquark
condensate, appears.
At further increase of $\mu$
the constituent quark mass decreases, and it would vanish at $\mu=\mu_1$,
\begin{equation}
\mu_1=\sqrt{\Lambda^2-\frac{\pi^2}{3G}}~,
\end{equation}
if the diquark condensate did not appear.
Actually, at the critical value $\mu_1$ both the quark condensate and
the diquark condensate are small but nonzero.

\begin{figure}
\includegraphics[scale=1.0]{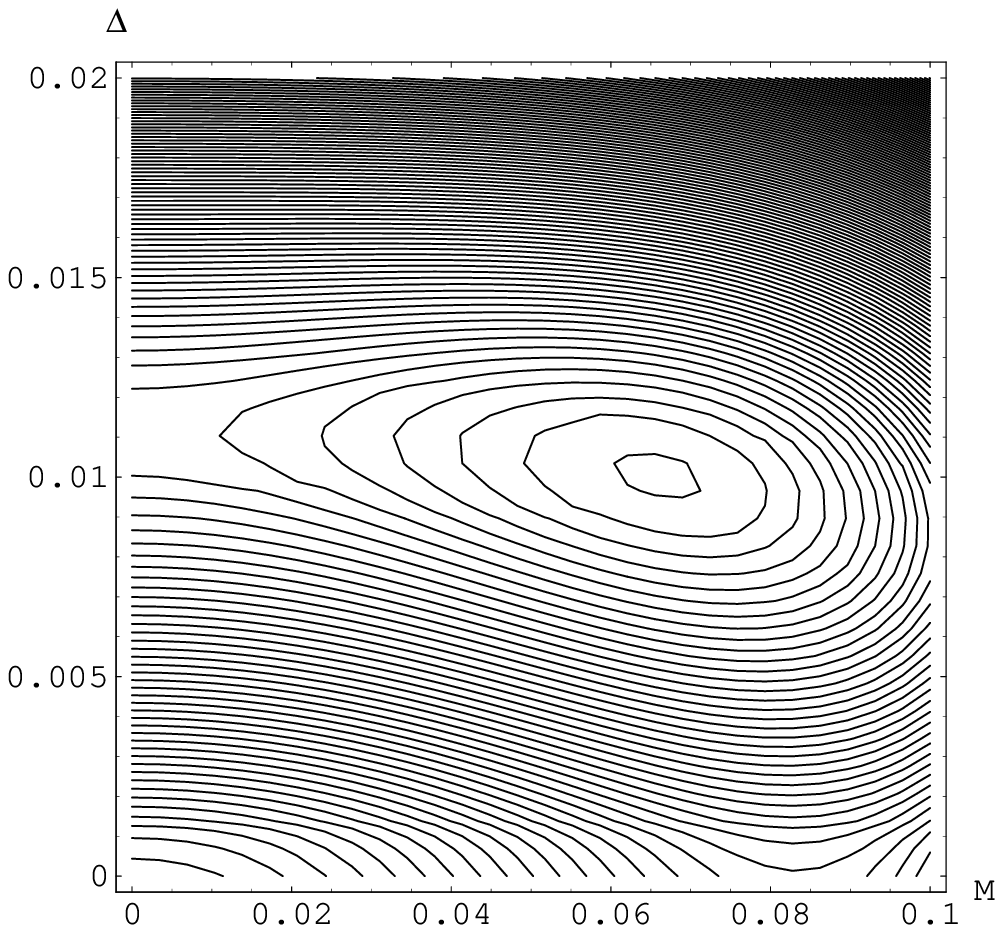}
\caption{The contour plot for the thermodynamic potential
as a function of of $m$ and $\Delta$ at zero temperature
and the chemical potential $\mu=$ 286 MeV for the Type II
parameter set.}
\label{thdp_II}
\end{figure}

The changes of the local extrema for increasing chemical potential
are similar to those shown in Figs.~\ref{thdp_I_1}--\ref{thdp_I_4}.
The cases of dilute ($\mu=0$) and very dense matter
($\mu=400$ MeV) are
qualitatively analogous,
only the absolute values of $m$ and $|\Delta|$
at which the local minima are found are different.
At intermediate densities, however, there is a qualitative
difference.
 Within a very narrow range of values of the chemical potential,
there exists a new phase of massive superconducting matter.
One can see this in Fig.~\ref{thdp_II}
for $\mu=286$ MeV.
At higher $\mu$ the chiral symmetry is restored and the
quark matter is in the pure superconducting phase.
A possibility of the chiral diquark condensates to coexist at certain condition
has been already noticed in Ref.~\cite{Sadzhi}

Thus, for the Type II parameter set, the transition from the
hadronic to the superconducting phase is of the second order.
In this case there are no degenerate local minima in the thermodynamical
potential separated by a barrier.
This behavior is unlike to what is commonly expected for a cold and  dense matter
but it parallels the findings of Ref.~\cite{kitazawa} where vector interactions are
responsible for this behavior.

\begin{figure}
\includegraphics[scale=0.7]{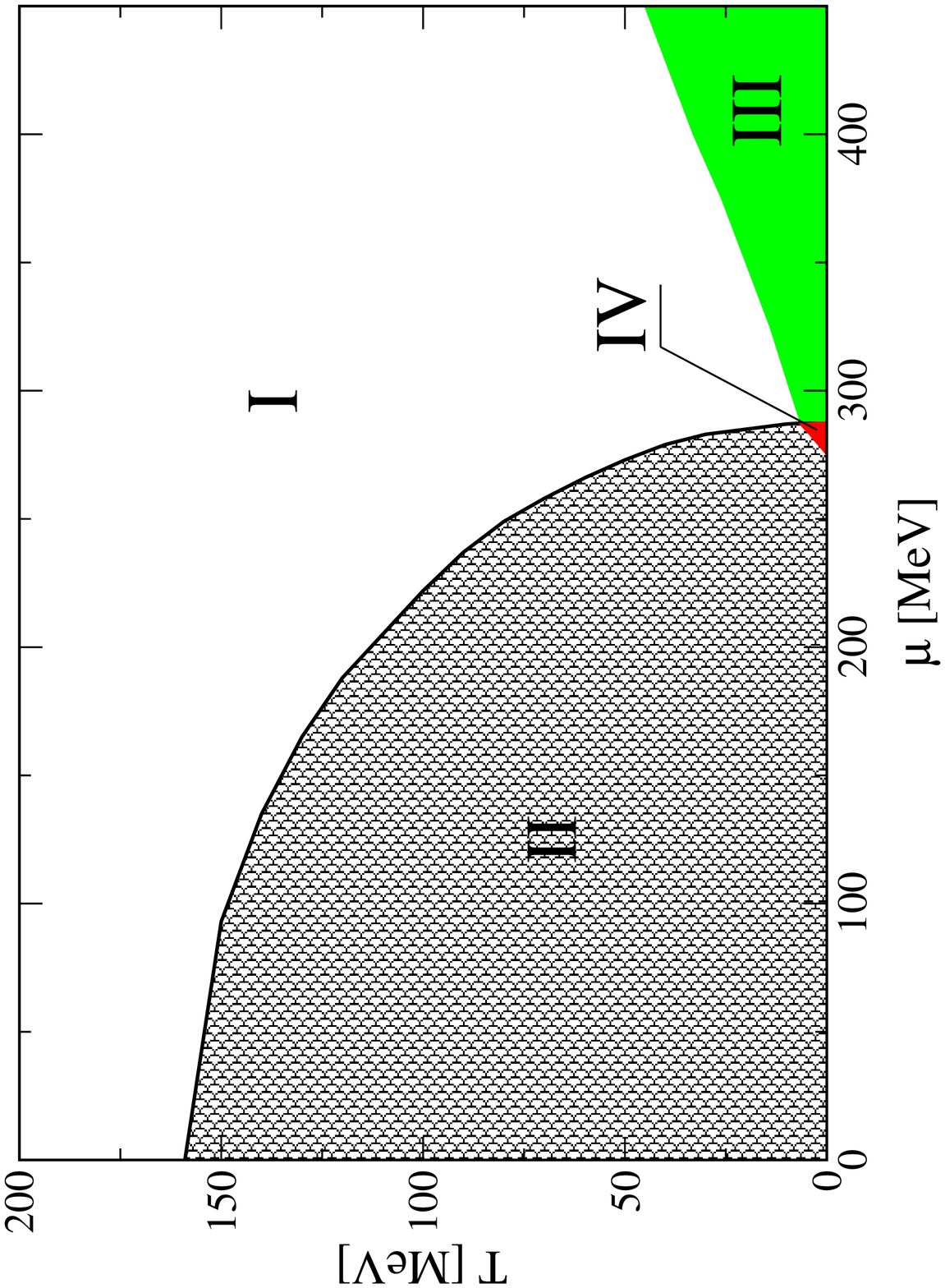}
\caption{The quark matter phase diagram from the NJL model
with the Type II parameter set. In phase I both chiral and diquark
condensates vanish; in phase II the chiral symmetry is broken; in
phase III the chiral symmetry is restored while the diquark condensate is nonzero;
in phase IV both the chiral and diquark condensates coexist.}
\label{phdII}
\end{figure}

The average value of $|\Delta|$ is much smaller than for
the Type~I parameter set.
As a consequence, the border between 2SC and the
symmetric phases of quark matter lies at noticeably  lower temperatures.
The phase diagram obtained in our model for the
Type~II parameter set is shown in Fig.~\ref{phdII}.

\section{Conclusion}
 In the framework of the simple NJL model for two flavors, a phase
diagram is obtained for $T=0$--200 MeV and $\mu=0$--450 MeV.
Three phases are found for the Type I parameter set and four phases
for the Type II parameter set. The critical temperature and
chemical potential  obtained in the Type I scheme differ from
those obtained with the Type II parameter set.
At $T=0$, $\mu_c\approx 320$ MeV for the Type I parameter set and
$\mu_c\approx 288$ MeV for the Type II.
The corresponding quark densities differ by a factor 1.5 -- 1.7.
The critical temperature for the Type II parameter
set is as low as $7$ MeV and thus much closer to critical temperatures for
the paring instability in nuclear matter systems  (see \cite{Sedrakian:1999cu})
whereas for the Type I parameter set the critical temperatures are an order of
magnitude larger.
This striking difference in the critical parameters obtained within the same
model calls for a more detailed investigation of the question of model
parametrization.

In our work, the constant $H$ was not obtained from a fit to observable data.
Instead, Fierz transformation arguments have been used to fix the ratio
$H/G=3/4$.
A parameterization would be favourable where (in the spirit of the Type II model)
experimentally measured quantities, like the $\rho$ meson width, are used rather
than non-observable ones (quark condensate etc.).
It would therefore be more consistent to fit the constant $H$ from baryon properties,
see \cite{pepin00,Bentz:2002um} and also to go beyond the mean field level of
description. These investigations shall be performed in future work where it remains
to be clarified which critical parameters for the color superconducting phase transition
can be considered more realistic and of which order the phase transition is.

\subsection*{Acknowledgement}
The authors are grateful to M. Buballa, T. Kunihiro, and M. Kitazawa for
useful  discussions.
This work has been supported by DAAD and Heisenberg-Landau programs.
V.Y. and M.K.V. acknowledge support by RFBR grant No. 02-02-16194.

\clearpage

\section*{Tables}

\begin{table}[h]
\begin{tabular}{||l|c|c|c|c|c|c||}
\hline\hline
			&$\Lambda$  & $G$           & $H$           & $m$       & $\sqrt[3]{-\vev{\bar\psi\psi}}$  & $g_\rho$\\
			& [GeV]     & [GeV$^{-2}$]  & [GeV$^{-2}$]  & [MeV]     & [MeV] & \\
\hline
GTR+QC      & 600       & 12.8          & 9.6           & 350       &    240       & 9.2 \\
{[Type I] }   & & & & & & \\
\hline
GTR+R2PD    & 856       & 5.1          & 3.8           & 233       &     284      & 6.1 \\
{[Type II]}   & & & & & &\\
 \hline\hline
\end{tabular}
\caption{The model parameters for two different schemes of parameter
fixing. The first row corresponds to the case where
the quark condensate value is used, while in the second row
the parameters correspond to the case where the decay width of the process $\rho\to2\pi$
is used.}
\label{paramset}
\end{table}
\clearpage

\section*{Figures}

\end{document}